\documentstyle[12pt]{article}
\begin{document}
\pagestyle{empty}
\vspace* {13mm}
\baselineskip=24pt
\begin{center}
{\bf GENERALIZED QUON STATISTICS}
\\[11mm]
Stjepan Meljanac and Ante Perica\\
{\it Rudjer Boskovic Institute, P.O.B. 1016,\\
41001 Zagreb, Croatia}\\[12mm]
{\bf Abstract}
\end{center}
\vskip 0.3cm
Generalized quons interpolating between Bose, Fermi, para-Bose,
para-Fermi, and anyonic statistics are proposed. They follow
from the R-matrix approach to deformed associative algebras.
It is proved that generalized quons have the same main properties
as quons. A new result for the number operator is presented and some physical
features of generalized quons are discussed in the limit $|q_{ij}^{2}|
\rightarrow 1$.
\newpage
\setcounter{page}{1}
\pagestyle{plain}
\def\leer{\vspace{5mm}}
\setcounter{equation}{0}
 Quons were proposed by Greenberg [1,2] as particles interpolating between
bosons and fermions. Quonic intermediate statistics is an example
of infinite statistics in which any representation of the
symmetric group can occur. It was pointed out that quons offered
a possibility for a small violation of the Pauli exclusion principle,
at least in nonrelativistic theory [2,3]. The quon algebra,
interpolating between bosonic and fermionic oscillators,
was postulated [2] as\\
\begin{equation}
a_{i}a_{j}^{\dagger} - qa_{j}^{\dagger}a_{i} = \delta_{ij},\ q\in R.
\end{equation}
Here i,j are discrete indices and the parameter $q$ is a real number.
 The main properties of quons are as follows:\\
 (i)  Norms are positive definite for $-1\leq q \leq 1$.\\
 (ii) For $q^{2}\neq 1$, the commutation relations do not exist
between annihilation (creation) operators $a_{i}$,$a_{j}$
($a_{i}^{\dagger}$,$a_{j}^{\dagger}$), i.e., there are $n!$ linearly
independent states $a_{i_{1}}\dots a_{i_{n}}|0>$ for different
permutations of fixed indices $1,2,\dots ,n$.\\
 (iii) The number operator exist in the form of an infinite
series expanded in powers of creation and annihilation operators,
with complicated coefficients diverging when $q^{2}\rightarrow 1$.\\
 (iv) The theory is nonlocal, but the TCP theorem and the clustering property
hold in relativistic quon theories.\\
 In a recent Letter [4] Mishra and Rajasekaran proposed a
q-deformed algebra of creation and annihilation operators with ordered
indices in which the deformation parameter was complex. The q-deformed algebra
was defined by the following equations:\\
\\
                $a_{i}a_{j}^{\dagger} - qa_{j}^{\dagger}a_{i} = 0$ for $i<j$\\
                $a_{i}a_{i}^{\dagger} - pa_{i}^{\dagger}a_{i} = 1$,\\
where $q$ and $p$ are complex and real parameters, respectively. Consequences
of this complex q-mutator algebra were studied and its relation to
"fractional" statistics was pointed out.\\
\ Anyons were proposed [5,6] as particles in $2+1$ dimension
that also interpolate between bosons and fermions.
It was shown [7] that multivalued anyons can be treated as a kind
of quons with the multivalued unimodular parameter $q = e^{i\lambda
\triangle}$, $\lambda $ being a real statistical parameter and
$\triangle = \pi +2\pi z$, $z\in Z$.\\
\ In this paper we propose generalized quons as particles interpolating
between all kinds of statistics simultaneously.
For a special choice of $q$ parameters
, one finds any particular statistics.
Generalized quons in principle allow for a small violation of any statistics.
We point out that,
generally, the $q$ parameter can depend on the pair of indices $i,j$
in the product $a_{i}a_{j}^{\dagger}$ in Eq.(1). Moreover, $q_{ij}$
can be complex numbers, with $q_{ij}^{*} = q_{ji}$. Hence the global parameter
$q$ becomes a Hermitian matrix. We point out that generalized quons follow
from the R-matrix approach to deformed algebras. We show that all properties
of quons hold for generalized quons. A new result for the number operator is
obtained. Some physical features of generalized quons are discussed in the
limit $|q_{ij}|^{2} \rightarrow 1$.\\
 We propose the generalized quon algebra as\\
 \begin{equation}
 a_{i}a_{j}^{\dagger} - q_{ij}a_{j}^{\dagger}a_{i} = \delta_{ij},
 \ q_{ij}^{*} = q_{ji},
 \ i,j\in S,
 \end{equation}

where $i,j$ denote sites of the discrete totally ordered lattice S in arbitrary
dimension.
This algebra interpolates between bosons (for $q=1$), fermions
(for $q=-1$), multivalued anyons [7] (for $q=e^{i \lambda \pi (1+2z)}$,
$z\in Z$), single-valued anyons [8,9] (for $q(\vec{r}-\vec{r^{'}}) =
\pm e^{i\lambda [\theta_{0}(\vec{r}-\vec{r^{'}})-\theta_{0}(\vec{r^{'}}
-\vec{r})]}$, where $\theta_{0}(\vec{r})$ is the polar angle and
+(-) corresponds to the Bose (Fermi)-type transformation). If $q_{ij}
=q$ for $i<j$, where $q$ is a complex parameter, and $q_{ii}=p$, where
$p$ is a real parameter, we obtain the q-deformed algebra of Ref.[4].\\
 In order to show that para-Bose and para-Fermi statistics [10] are also
contained in the general quon algebra, Eq.(2), let us consider a set
of operators $a_{i}^{\alpha}$, $\alpha = 1, \dots ,p$, satisfying the
algebra \\
\begin{equation}
a_{i}^{\alpha}a_{j}^{\beta \dagger} - q_{\alpha \beta}a_{j}^{\beta \dagger}
a_{i}^{\alpha} = \delta_{\alpha \beta}\delta_{ij}, \ q_{\alpha \beta}^{*}
= q_{\beta \alpha}.
\end{equation}

Then $q_{\alpha \beta} = \pm 1$ for $\alpha = \beta $ and $q_{\alpha \beta}
= \mp 1$ for $\alpha \neq \beta $ correspond to Green's para-Bose (para
-Fermi) type of oscillators for the upper (lower) sign.\\
 We point out that the generalized-quon algebra, Eq.(2), is just a special
type of associative algebra; namely, the general commutation algebra
of oscillators can be written in the R-matrix approach [11,12]:\\
\begin{equation}
\begin{array}{l}
a_{i}a_{j} - R_{ij,kl}a_{l}a_{k} = 0,\\
a_{i}a_{j}^{\dagger} - R_{ki,jl}^{'}a_{k}^{\dagger}a_{l} = Q_{ij}.
\end{array}
\end{equation}

Here $R$ and $R^{'}$ are matrices with complex entries, whereas
$(Q_{ij})$ is a set of operators with $Q_{ij}^{\dagger} = Q_{ji}$.
In the following we restrict ourselves to the case $Q_{ij} = \delta_{ij}$.\\
 In order that the algebra in Eq.(4) be associative, the following condition
have to be satisfied:\\
\ (a) the Yang-Baxter equation (summation over repeated indices is assumed) \\
  \begin{equation}
 R_{jk,xy}R_{iy,zb}R_{zx,am} = R_{ij,xy}R_{xk,az}R_{yz,mb},
 \end{equation}

 (b) \\
 \begin{equation}
 R_{jk,xy}R_{iy,zb}^{'}R_{zx,am}^{'} = R_{ij,xy}^{'}R_{xk,az}^{'}R_{yz,mb},
 \end{equation}

 (c) the Hecke condition, symbolicaly written as\\
 \begin{equation}
 (\hat{R} - 1)(\hat{R^{'}} +1) = 0,
 \end{equation}

where $\hat{R}=PR$, $\hat{R^{'}}=PR^{'}$ and $P$ is the permutation operator
$P_{ij,kl}=\delta_{il}\delta_{jk}$,\\
\ (d) hermiticity, i.e., that $a_{i}^{\dagger}$ is the hermitian conjugate of
 $a_{i}$ (and vice versa):\\
 \begin{equation}
 R_{ij,kl}^{'} = R_{lk,ji}^{'*},\ i.e., \ (\hat{R^{'}})^{\dagger} =
 \hat{R^{'}}.
 \end{equation}

Solutions of Eq.(5-8), corresponding to Bose, para-Bose, Fermi, para-Fermi,
and anyonic statistics are \\
 $R =R^{'} = \pm 1$, upper (lower) sign for bosons (fermions),\\
 $R = R^{'} = e^{i\lambda \pi (1+2z)}\cdot 1$, $z\in Z$, for multivalued
 anyons [7],\\
 $R(\vec{r_{1}},\vec{r_{2}},\vec{r_{3}},\vec{r_{4}}) = R^{'}(\vec{r_{1}},
 \vec{r_{2}},\vec{r_{3}},\vec{r_{4}})$\\$ = \pm e^{i\lambda [\theta_{0}
 (\vec{r_{1}}-\vec{r_{2}})-\theta_{0}(\vec{r_{2}}-\vec{r_{1}})]}
 \delta (\vec{r_{1}}-\vec{r_{3}})\delta (\vec{r_{2}}-\vec{r_{4}})$\\
 for single-valued anyons [8,9] with +(-) for the Bose (Fermi)-type
 transformation, and\\
 $R = R^{'}$,\ $R_{\alpha i,\beta j,\gamma k,\delta l} =
 q_{\alpha \beta}\delta_{\alpha \gamma}\delta_{\beta \delta}\delta_{ik}
 \delta_{jl}$,\\
 where $\alpha ,\beta ,\gamma ,\delta = 1, \dots ,p$ and $q_{\alpha \beta}
 = \pm (2\delta_{\alpha \beta} - 1)$, +(-) corresponding to Green's
 oscillators of the para-Bose (para-Fermi) type [10].\\
 However, there is also another solution of Eqs.(5-8), $R = P$,i.e.,
$\hat{R} = 1$, and an arbitrary Hermitian matrix $R^{'}$.
Specially, we put \\
\begin{equation}
(R^{'})_{ki,jl} = q_{ij}\delta_{jk}\delta_{il}, \ q_{ij}=q_{ji}^{*}.
\end{equation}

 The corresponding associative algebra is just the generalized-quon
algebra in Eq.(2). There are no relation between $a_{i},a_{j}$
(or $a_{i}^{\dagger},a_{j}^{\dagger}$) operators.
When $R^{'}$ approaches one of the above special solutions $R=R^{'}$,
quons approach bosons, fermions, anyons, etc., allowing for a small
violation of the corresponding approaching statistics. In the exact limit,
the relations between $a_{i},a_{j}$ (or $a_{i}^{\dagger},a_{j}^{\dagger}$)
appear.\\
 Finally, we have proved that all the main properties of quons still
hold for generalized quons.\\
 (i) The norms are positive definite for $|q_{ij}| < 1$.\\
We assume the existence of a vacuum state $|0>$ and its dual $<0|$
satisfying \\
\begin{equation}
a_{i}|0> = 0,\ \ \ <0|a_{i}^{\dagger} = 0 ,\ \forall i \in S.
\end{equation}\\

The Fock-like space is constructed in the usual way. The one-particle states
are $a_{i}^{\dagger}|0>$, $i\in S$, and, generally, the n-particle states are
of the
form $a_{i_{1}}^{\dagger}\cdots a_{i_{n}}^{\dagger}|0>$, $i_{1},\dots
,i_{n} \in S$. When the indices $i_{1},\dots ,i_{n}$ are mutually different
there may exist maximally $n!$ linearly independent states specified by
permutations of $i_{1},\dots ,i_{n}$. If $n_{1},\dots ,n_{a}$
are multiplicities of equal indices appearing in an ordered sequence
$i_{1},\dots ,i_{n}$ ($i_{1}\leq i_{2}\leq \cdots \leq i_{n}$) satisfying
$\sum_{\alpha =1}^{a}n_{\alpha} =n$, then there may exist maximally
$\frac{n!}{n_{1}!\cdots n_{a}!}$ linearly independent states specified
by permutations of $i_{1},\cdots i_{n}$. The dual (bra) states are defined
by $<0|a_{i_{n}}\cdots a_{i_{1}}$,\  $i_{1},\dots i_{n}\in S$.\\
Let us define the
matrix $A$ of inner products with the matrix elements :\\
\begin{equation}
A_{i_{1},\dots ,i_{m};j_{1},\dots ,j_{n}} =
<0|a_{i_{m}}\cdots a_{i_{1}}a_{j_{1}}^{\dagger}\cdots a_{j_{n}}^{\dagger}
|0>.
\end{equation}

This is in fact the vacuum matrix element of any polynomial $a_{i_{m}}\cdots
a_{i_{1}}a_{j_{1}}^{\dagger}\cdots a_{j_{n}}$, that can be calculated using
Eqs(2),(10). The matrix $A$ is hermitian and block-diagonal. If the lattice
S is finite with number of sites D, then
there are
\[ \left( \begin{array}{c}
D+n-1 \\
n
\end{array} \right) \]
 different blocks (of size $\leq n!$) in the n-particle sector. A generic block
$A^{(i_{1}\cdots i_{n})}$ is characterizied
by mutually different ordered indices $i_{1},\dots i_{n}\in S$ $(i_{1}<
i_{2}< \cdots < i_{n}), n\leq D$, from which all other blocks in the
n-particle sector can be obtained using a suitable specification. The
$A^{(i_{1},\dots
i_{n})}$ matrix is an
$n!\cdot n!$ matrix, whose diagonal matrix elements are equal to $1$.
The arbitrary matrix element $(\pi ,\sigma)$, i.e.,$i_{\pi (1)},\cdots
i_{\pi (n)}; i_{\sigma (1)}\cdots i_{\sigma (n)}$, where $\pi $ and
$\sigma $ are permutations acting on positions $1,2\dots n$
($\pi $ denotes the row and $\sigma $ the column of the matrix $A^{(i_{1},\dots
i_{n})}$)
is given by \\

\begin{equation}
A^{(i_{1},\dots ,i_{n})}_{\pi ,\sigma } =
\prod_{\alpha, \beta} q_{i_{\alpha }i_{\beta }}.
\end{equation}

Here the product is over all pairs $\alpha ,\beta =1,\dots ,n$
satisfying $\pi^{-1}(\alpha) < \pi^{-1}(\beta)$ and
$\sigma^{-1}(\alpha) > \sigma^{-1}(\beta)$.
For $q_{ij}=q \in \cal{R}$, $\forall i,j\in S$, Eq.(12) reproduces the result
$A^{(i_{1},\dots ,i_{n})}_{\pi ,\sigma} = q^{I(\sigma^{-1}\cdot \pi)}$,
where $I$ denotes the number of inversions in permutation $\sigma^{-1}
\cdot \pi $, [2,13].\\

 For example, for $n=2$, Eq.(12) gives \\
\begin{equation}
A^{(i_{1},i_{2})} = \left(\begin{array}{cc}
                1 & q_{i_{1}i_{2}} \\
                q_{i_{2}i_{1}} & 1
                \end{array} \right) , i_{1},i_{2}\in S, i_{1}\neq i_{2}.
\end{equation}\\
Note that if $i_{1}=i_{2}=i$ the block matrix reduces to the one-by-one
matrix $A^{(i,i)}=1+q_{ii}$, $i\in S$.\\
\ Now we analyze the positivity for the norm (of all vectors) in the
Fock-like space. The norm of any one-particle state is positive since
$<0|a_{i}a_{i}^{\dagger}|0> = 1$, $\forall i\in S$.\\ Let us consider
two-particle states $(\alpha a_{i_{1}}^{\dagger}a_{i_{2}}^{\dagger}
+ \beta a_{i_{2}}^{\dagger}a_{i_{1}}^{\dagger})|0>$, $\alpha , \beta
\in \cal{C}$, $i_{1},i_{2}\in S$.\\
If $i_{1}\neq i_{2}$, the norm is \\
\begin{equation}
|\alpha |^{2} + |\beta|^{2} +\alpha ^{*}\beta q_{i_{1}i_{2}}+
\alpha \beta^{*}q_{i_{2}i_{1}}
\end{equation}\\

Hence the norms of all two-particle states will be positive if and only
if the matrices $A^{(i_{1},i_{2})}$ are positive definite for
$\forall i_{1},i_{2} \in S$. These conditions are (see Eq(13))\\
\begin{equation}
1 - |q_{i_{1}i_{2}}|^{2} > 0,\ \  i.e.\ \  |q_{i_{1}i_{2}}| < 1 \ \ i_{1}\neq
i_{2};\ i_{1},i_{2} \in S.
\end{equation} \\
When $i_{1}=i_{2}=i$, the norms are positive if \\
\begin{equation}
A^{(i,i)} = 1 + q_{ii} > 0, \ i.e.,\ \ q_{ii}>-1,\ \ \forall i \in S.
\end{equation}
\\
Note that when the inequalities (15) are satisfied,
the two particle states $a_{i_{1}}^{\dagger}a_{i_{2}}^{\dagger}|0>$,
and $a_{i_{2}}^{\dagger}a_{i_{1}}^{\dagger}|0>$ are linearly independent.
The norms of symmetric and antisymmetric two-particle states $||\frac{1}{2}(
a_{i_{1}}^{\dagger}a_{i_{2}}^{\dagger} \pm
a_{i_{2}}^{\dagger}a_{i_{1}}^{\dagger}|0>||^{2}$ are $\frac{1}{2}
(1 \pm Re q_{i_{1}i_{2}})$, where the upper (lower) sign corresponds to the
symmetric (antisymmetric) state. However, these states are not orthogonal
if $Im q_{i_{1}i_{2}}\neq 0$. The eigenstates of $A^{(i_{1},i_{2})}$
are $\frac{1}{2}(a_{i_{1}}^{\dagger}a_{i_{2}}^{\dagger}
 \pm e^{-i\phi}a_{i_{2}}^{\dagger}
a_{i_{1}}^{\dagger})|0>$, $\phi = Im q_{i_{1}i_{2}}$, and they are
orthogonal. The
occupation probabilities that the state
$a_{i_{1}}^{\dagger}a_{i_{2}}^{\dagger}|0>$ is in these eigenstates are
$\frac{1}{2}(1 \pm |q_{i_{1}i_{2}}|)$, respectively.\\
Proceeding in this way we demand that every n-particle state should have
positive
norm. We find that the norms of all n-particle states will be positive if
and only if
the matrices $A^{(i_{1},\dots ,i_{n})}$ are positive definite for every
ordered sequence of indices $i_{1},\dots ,i_{n}$, $i_{1}\leq i_{2}\leq \cdots
\leq
i_{n}$.\\
\ Furthermore, we point out that for $q_{ij}=0$, $\forall i,j\in S$, all
conditions for positive definiteness are satisfied automatically, since
$A$ is the identity matrix.
There arises a natural question how far one can change $q_{ij}$, $\forall
i,j\in S$, starting from zero in order
that all eigenvalues remain positive, i.e., all matrices
$A^{(i_{1},\dots ,i_{n})}$ remain positive definite.
This will happen as long as $det A^{(i_{1},\dots ,i_{n})} > 0$ for
$\forall i_{1},\dots ,i_{n}\in S$, $i_{1}\leq i_{2}\leq \cdots \leq i_{n}$
since the eigenvalues of $A^{(i_{1},\dots ,i_{n})}$ are real and depend
continuously on $q_{ij}$, $\forall i,j\in S$.

The determinant of a $n! \cdot n!$ generic matrix $A^{(i_{1},\dots
,i_{n})}$, $i_{1}<i_{2}<\cdots <i_{n}$, is given by [14] \\

\begin{equation}
det A^{(i_{1},\dots ,i_{n})} = \prod^{n-1}_{k=1}\{ \prod_{(j_{1},\dots
,j_{k+1})}
[1 -\prod_{\alpha}\prod_{\beta}|q_{j_{\alpha}j_{\beta}}|^{2}]\} ^{(k-1)!
(n-k)!}
\end{equation}

where $1 \leq \alpha < \beta \leq k+1$, and the second product is over
\[ \left(  \begin{array}{c}
n \\
k+1
\end{array} \right) \]
combinations of indices $(j_{1},\dots ,j_{k+1})\subset (i_{1},
\dots ,i_{n})$. When $q_{ij}=q\in \cal{R}$, $\forall i,j\in S$, Eq.(17)
reproduces the result of Zagier
[13]. We have not found a general expresion  for the determinant
of the reduced matrix $A^{(i_{1},\dots ,i_{n})}$ when some of the indices
coincide. However, the determinant of the reduced matrix $A^{(i_{1},\dots
,i_{n})}$ when some indices coincide is a polynomial factor of the
determinant of the generic matrix, Eq.(17), after identifying
the corresponding indices [14]. Hence, a sufficient condition
for the positivity of all norms is that the expression in Eq.(17)
should be positive for every ordered sequence $i_{1},\dots ,i_{n}\in S$
including repetitions and for $\forall n$. It follows that the
Fock-like space is positive definite for $|q_{ij}| < 1$, $\forall
i,j \in S$. \\
Note that in the three-particle sector the positivity of the reduced matrix
$A^{(iij)}$, when two of the indices coincide, implies
$-1 < q_{ii} < |q_{ij}|^{-2}$, $\forall j\in S$, $j\neq i$.
Hence, in some cases, one might expect that $q_{ii}$ could be larger
than 1.\\

 (ii) The number operator $N_{k}$ exists for generalized quons, satisfying
$[N_{k},a_{l}] = -a_{k}\delta_{kl}$. We present a new result for the number
operator [15] that exhibits its simple structure and is given by \\
\begin{equation}
N_{k} = a_{k}^{\dagger}a_{k} + \sum^{\infty}_{n=1}\sum_{i_{1},\dots ,i_{n}}
\sum_{\pi \in S_{n}}[Y_{k,\pi(i_{1},\dots ,i_{n})}]^{\dagger}
Y_{k,i_{1},\dots ,i_{n}}(A^{(k,i_{1},\dots ,i_{n})})^{-1}_{k,i_{1},\dots
,i_{n};
k,\pi(i_{1},\dots ,i_{n})} ,
\end{equation}

where $S_{n}$ denotes the permutation group, and \\
\begin{equation}
Y_{k,i_{1},\dots ,i_{n}} = Y_{k,i_{1},\dots ,i_{n-1}}a_{i_{n}} -
q_{i_{n}k}q_{i_{n}i_{1}}\cdots q_{i_{n}i_{n-1}}a_{i_{n}}
Y_{k,i_{1},\dots ,i_{n-1}}
\end{equation}

and \\
\begin{equation}
Y_{ki_{1}} = a_{k}a_{i_{1}} - q_{i_{1}k}a_{i_{1}}a_{k}.
\end{equation}

 Always when the parameters $q_{ij}$ tend to $1$, i.e., $q_{ij}
\rightarrow 1$, for $\forall i,j$, quons tend to a particular anyonic-type
statistics, the matrices $(A^{(i_{1},\dots i_{n})})^{-1}$ become singular and
the coefficients
in the number operator, Eq.(18), diverge. Nevertheless, the number operator
$N_{k}$, when acting on states, is well defined. Moreover, in the exact limit
it reduces to $N_{k} = a_{k}^{\dagger}a_{k}$, and additional relations between
the annihilation (creation) operators $a_{i},a_{j}$ ($a_{i}^{\dagger},a_{j}
^{\dagger}$) emerge. In this case, the corresponding particles are not
distinguishable, i.e., they are identical in the quantum-mechanical sense.
Interchanging them, we generally obtain a unit phase factor $e^{i\alpha}$
(typical of anyons). The n-particle wave functions $a_{i_{1}}^{\dagger}
\cdots a_{i_{n}}^{\dagger}|0>$ for different permutations of indices are
linearly dependent. Symmetries of these states can be classified
according to the q-symmetrization or q-antisymmetrization prescription.
Otherwise, for general quons there are $n!$ linearly independent states
$a_{i_{1}}^{\dagger}\cdots a_{i_{n}}^{\dagger}|0>$, for different
permutations of indices $1,\dots ,n$.\\
\ (iii) The theory of generalized quons is nonlocal. The TCP theorem
and the clustering properties still hold, with the same type of arguments
as in Ref.1 and 2. However, a careful relativistic treatment of generalized
quons should be undertaken, see the discussion in Refs.16 and 17.\\

 We thank Dr. M. Milekovic and Prof. D. Svrtan for helpful discussions.
This work was supported by the Scientific Fund of the Republic of Croatia.

%

\end{document}